\documentclass[aps,pra,onecolumn,showpacs,superscriptaddress,groupedaddress]{revtex4}  
\usepackage{graphicx}  
\usepackage{dcolumn}   
\usepackage{bm}        
\usepackage{amssymb}   
\usepackage[colorlinks = true,
            linkcolor = blue,
            urlcolor  = blue,
            citecolor = blue,
            anchorcolor = blue]{hyperref}
\begin{document}
	
	
	
\title{Simultaneous dual-species laser cooling using an optical frequency comb}
	
\author{D.~Buhin}
\affiliation{Institute of Physics, Bijeni\v{c}ka cesta 46, 10000 Zagreb, Croatia}
\author{D.~Kova\v{c}i\'{c}}
\affiliation{Institute of Physics, Bijeni\v{c}ka cesta 46, 10000 Zagreb, Croatia}
\author{F.~Schmid}
\affiliation{Max-Planck-Institut f\"{u}r Quantenoptik, 85748 Garching, Germany}
\author{M.~Kruljac}
\affiliation{Institute of Physics, Bijeni\v{c}ka cesta 46, 10000 Zagreb, Croatia}
\author{V.~Vuli\'{c}}
\affiliation{Institute of Physics, Bijeni\v{c}ka cesta 46, 10000 Zagreb, Croatia}
\author{T.~Ban}
\affiliation{Institute of Physics, Bijeni\v{c}ka cesta 46, 10000 Zagreb, Croatia}
\author{D.~Aumiler}
\affiliation{Institute of Physics, Bijeni\v{c}ka cesta 46, 10000 Zagreb, Croatia}
\email{aumiler@ifs.hr}

\date{\today}

\begin{abstract}
We demonstrate 1D simultaneous laser cooling of $^{87}$Rb and $^{85}$Rb atoms using an optical frequency comb.
By adjusting the pulse repetition frequency and the offset frequency, the frequency comb spectrum is tuned to ensure that two distinct frequency comb modes are simultaneously red-detuned from the cooling transitions, one mode for each species. 
Starting from a pre-cooled cloud of $^{85,87}$Rb atoms at above-Doppler temperatures, we show simultaneous cooling of both species down to the Doppler temperature using two counter-propagating $\sigma$$^{+}$/$\sigma$$^{-}$-polarized beams from the frequency comb. 
The results indicate that simultaneous dual-species frequency comb cooling does not affect the cooling characteristics of individual atomic species. 
The results of this work imply that several atomic species could be cooled simultaneously using a single frequency comb source. 
This comb-based multi-channel laser cooling could bring significant advances in multi-species atom interferometers for space applications and in the study of multi-species interactions.
		
\end{abstract}
	
\maketitle
	
Optical frequency combs (FCs) are unique light sources with applications ranging from metrology \cite{ye2003} and high-resolution spectroscopy \cite{maslowski2014, picque2019} to quantum communication and processing \cite{maltese2020}. 
In the time domain, the output of a FC is a train of phase-stable ultrashort pulses of typically high peak power. 
This allows for efficient frequency conversion and other non-linear interactions.
The spectrum of a FC consists of a series of equally spaced narrow spectral lines.
Owing to these unique properties, FCs have been proposed as potential light sources for laser cooling of atoms with strong cycling transitions in the vacuum ultraviolet (VUV) \cite{kielpinski2006, udem2016, cambell2016}.
This part of the spectrum, and thus some of the most prevalent atomic species, has so far remained inaccessible to laser cooling since generating continuous wave (CW) laser radiation in the VUV is extremely challenging.
	
Cooling of atoms and ions using FCs has recently been demonstrated. 
Jayich et al. \cite{cambell2016} achieved FC Doppler cooling of pre-cooled rubidium atoms on the two-photon transition at 778 nm. 
Davila-Rodriguez et al. \cite{udem2016} showed FC Doppler cooling of trapped magnesium ions on a single-photon transition in the UV.
Ip et al. \cite{ip2018} demonstrated loading, cooling and crystallization of hot ytterbium ions, and \v{S}anti\'{c} et al. \cite{santic2019} demonstrated cooling of rubidium atoms on a single-photon transition at 780 nm.
A recent theoretical work by our group \cite{aumiler2012} investigated simultaneous laser cooling in multiple cooling channels using a FC and has shown that simultaneous cooling of $^{40}$K, $^{85}$Rb, and $^{87}$Rb can be achieved using a single FC by appropriate selection of comb parameters, i.e. pulse repetition frequency and offset frequency.
	
In this Rapid Communication, we demonstrate simultaneous laser cooling of $^{85}$Rb and $^{87}$Rb atoms using an optical frequency comb. 
To our knowledge, this is the first demonstration of FC cooling in multiple cooling channels simultaneously.
We believe that the application of FC multi-channel cooling could bring significant advances in multi-species atom interferometers (AIs).
Simultaneous dual-species AIs pave the way for future ground and space experiments dedicated to testing the weak equivalence principle, also known as the universality of free fall \cite{bonnin2013, zhou2015, williams2016}.
Current design for space applications is based on the $^{85}$Rb/$^{87}$Rb dual-species interferometer which employs four amplified diode laser modules at 780 nm, offset locked to the rubidium spectroscopy referenced frequency doubled Telecom laser, for the simultaneous laser cooling and coherent manipulation of atoms \cite{schuldt2015}.
Multi-species AIs offer extended dynamic measurement ranges \cite{bonnin2018} which could increase the sensitivity and resolution of the instruments.
To our knowledge, three- (and more than three) species AIs have so far not been demonstrated, most likely due to the complexity of the laser systems required.
In this context, we believe that the application of frequency combs with multi-species cooling capabilities, potentially utilizing the developing chip-based microresonator FC technology \cite{kippenberg2011, tanabe2019, savchenkov2013}, could lead to a breakthrough in the development of multi-species AIs for space applications.
	
Dual- and multi- species magneto-optical traps (MOTs) are an experimental tool for investigating atomic interactions. 
They are a starting point for the production of quantum degenerate mixtures \cite{modugno2002, hara2011,vaidya2015, madison2013} as well as for the formation of heteronuclear cold molecules \cite{carr2009, menegatti2008}. 
FC multi-channel cooling demonstrated in this work could greatly reduce the complexity of multi-species MOT experimental systems by replacing a series of CW lasers (independent cooling and repumper lasers are usually required for each species) with a single frequency comb source where different modes within the comb spectrum can serve as cooling as well as repumper lasers.
One notable simplification of the experiment involves replacing a large number of feedback loops (one feedback loop is required to stabilize each individual CW laser) with only two feedback loops that can stabilize and phase-lock all lines within the FC spectrum.
FC cooling would therefore allow cooling of different atomic species by highly phase-coherent frequency comb modes which could bring new insights into the physics of heteronuclear cold collisions and molecules formation \cite{aumiler2012, chin2010}. 
	
Our apparatus consists of a dual-species MOT in which $\approx 1\cdot10^6$ $^{85}$Rb atoms and $\approx 3\cdot10^6$ $^{87}$Rb atoms are simultaneously loaded from a background vapor in a stainless steel chamber.
The MOT relies on four independent frequency-stabilized CW lasers arranged in a standard six-beam configuration, which together with a quadrupole magnetic field creates the trapping potential for both species.
To ensure that the $^{85}$Rb and $^{87}$Rb MOTs are well overlapped, the cooling beams for both species are delivered through a single optical fiber. 
A detailed description of the experimental setup and the loading characteristics of the dual-species MOT can be found in the Supplemental Material.
In typical experimental conditions, we obtain two clouds of cold $^{85}$Rb and $^{87}$Rb atoms with temperatures in the range of 200-300 $\mu$K.
The clouds typically have 1/e$^{2}$ radii of $\approx $ 0.8 mm, and their centers of mass overlap to within 5 $\%$ of their radii.
Such overlapped pre-cooled clouds of $^{85}$Rb and of $^{87}$Rb atoms represent the initial sample for all measurements presented in this work.
	
The FC is generated by frequency doubling an Er:fiber mode-locked laser (TOPTICA FFS) operating at 1550~nm with a nominal repetition rate $f_{\text{rep}}=80.5$~MHz. 
A repetition rate tuning range of 50~kHz can be achieved by adjusting the laser cavity length with the help of an integrated stepper motor and a piezo transducer.
The frequency-doubled spectrum used in the experiment is centered around 780 nm with a FWHM of about 5~nm and a total power of 68~mW. 
The spectrum of the FC consists of a series of sharp lines, i.e. comb modes \cite{cundiff2005}. 
The optical frequency of the $n$-th comb mode is given by $f_n=n \cdot f_{\text{rep}}+f_0$, where $f_{\text{rep}}$ is the laser repetition rate and $f_0$ is the offset frequency.

In our experiment, we actively stabilize $f_{\text{rep}}$ and $f_n$ by giving feedback to the cavity length and pump power of the mode-locked laser, thus indirectly fixing $f_0$.
The $n$-th comb mode, $f_n$, is phase-locked to a frequency-shifted CW reference laser (ECDL, Moglabs CEL002), which is locked to the $^{87}$Rb $|5S_{1/2};F=2\rangle \rightarrow |5P_{3/2};F'=3\rangle$ transition.
The frequency shift of the CW reference laser is achieved by an acousto-optic modulator (AOM) in a double pass configuration.
$f_{\text{rep}}$ is stabilized to a low-noise synthesizer which is referenced to a rubidium frequency standard.
A detailed description of the FC stabilization scheme is presented in our recent paper \cite{santic2019}.
	
In order to achieve simultaneous cooling of two atomic species, two distinct modes within the comb spectrum must be simultaneously red detuned from the cooling transitions of the atomic species \cite{aumiler2012}.
Careful tailoring of the FC spectrum, i.e. choosing the appropriate $f_{\text{rep}}$ and $f_0$, is therefore crucial for the successful realization of FC cooling.  
In our experiment, the repetition rate is fixed during the measurements and set to $f_{\text{rep}}=80.495$~MHz, while $f_0$ is scanned by adjusting the heterodyne beat frequency between the CW reference laser and the $n$-th comb mode, $f_n$.
This way it is possible to control the detuning of the $n$-th comb mode with respect to the $^{87}$Rb $|5S_{1/2};F=2\rangle \rightarrow |5P_{3/2};F'=3\rangle$ transition.   
The heterodyne beat frequency can be continuously changed over a range of 5-30 MHz by changing the frequency of the local oscillator, so four separate scans with different CW reference laser frequency shifts are performed and subsequently merged to fully scan the $n$-th comb mode frequency by one $f_{\text{rep}}$. 

We start the investigation of simultaneous interaction of the FC with cold $^{85}$Rb and $^{87}$Rb atoms by measuring the FC radiation pressure force.
The experimental setup and the measurement sequence used are similar to the ones described in our recent work \cite{santic2019}.
A single circularly polarized FC beam is sent through an AOM for fast switching, and is then directed to the center of the dual-species MOT.
The total power of the FC beam on the atoms is 10~mW and the beam size (FWHM) is 2.7~mm, resulting in the power and intensity per comb mode of about 0.3 $\mu$W and 3.6 $\mu$W/cm$^{2}$, respectively.
	
\begin{figure}[h]
		\centerline{
			\mbox{\includegraphics[width=0.7\textwidth]{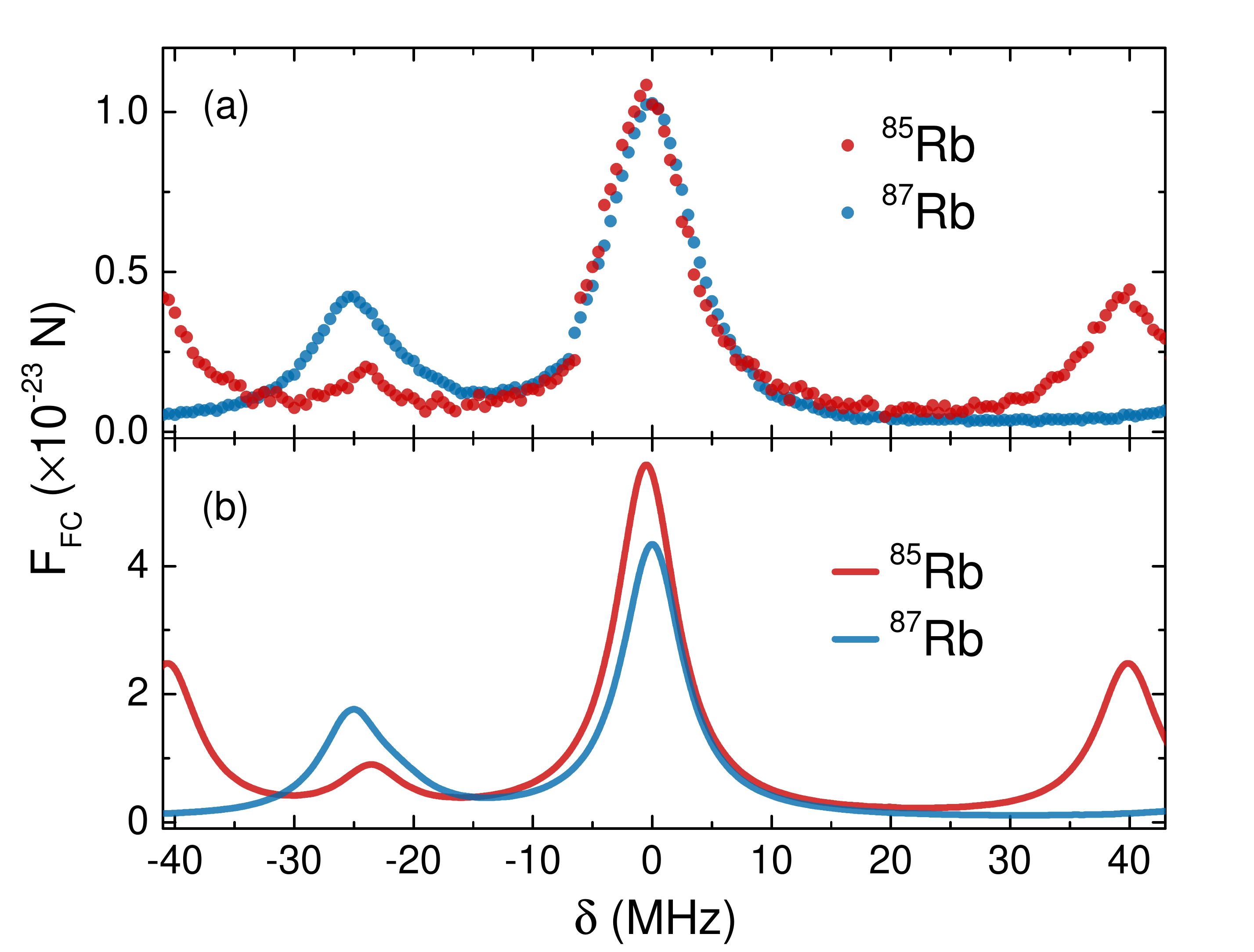}}
		}
		\caption{Measured (a) and calculated (b) FC radiation pressure force as a function of comb detuning $\delta$ for $^{85}$Rb (red) and $^{87}$Rb (blue) atoms. $\delta$ denotes the detuning of the $n$-th comb mode from the $^{87}$Rb $|5S_{1/2};F=2\rangle \rightarrow |5P_{3/2};F'=3\rangle$ transition. The calculated FC force in (b) includes contributions from three $^{85}$Rb $|5S_{1/2};F=3\rangle \rightarrow |5P_{3/2};F'=2,3,4\rangle$ and three $^{87}$Rb $|5S_{1/2};F=2\rangle \rightarrow |5P_{3/2};F'=1,2,3\rangle$ hyperfine transitions. The force contributions arising from each hyperfine transition are separately shown in the Supplemental Material.    
		}
		\label{Fig1}
\end{figure}

The measurement sequence starts with the preparation of cold $^{85}$Rb and $^{87}$Rb clouds and the FC beam off.
At $t=0$ we turn off the MOT cooling beams and switch on the FC beam.
The MOT repumper lasers are left on continuously to optically pump the atoms out of the $^{85}$Rb $|5S_{1/2};F=2\rangle$ and $^{87}$Rb $|5S_{1/2};F=1\rangle$ ground states.
They are arranged in a counter-propagating configuration with the intensity predominantly in the direction perpendicular to the FC beam propagation, and have no measurable mechanical effect (see Supplemental Material for a detailed scheme of the optical setup).
The quadrupole magnetic field is also left on.
We let the comb interact with the cold clouds for 2 ms.
During this time the centers of mass of both clouds accelerate in the FC beam direction ($+x$-direction) due to the FC radiation pressure force.
The FC beam and repumper lasers are then switched off and the clouds expand freely for 2 ms, after which we switch on the MOT cooling beams for 0.15~ms and image the cloud fluorescence with a camera. 
Fluorescence images are recorded separately for the $^{85}$Rb and $^{87}$Rb atoms.
This basic measurement sequence is repeated 6 times, and the resulting fluorescence images are then averaged for each isotope.
The cloud center of mass displacement in the $+x$-direction is determined from the images (for each isotope), providing information on the cloud acceleration and the FC radiation pressure force.
In Fig. \ref{Fig1} we show the measured (a) and calculated (b) FC force for each rubidium isotope as a function of the FC detuning $\delta$, which is defined as the detuning of the $n$-th comb mode from the  $^{87}$Rb $|5S_{1/2};F=2\rangle \rightarrow |5P_{3/2};F'=3\rangle$ transition. 
Due to the nature of the comb spectrum, the FC radiation pressure force is periodic with respect to comb detuning with period equal to $f_{\text{rep}}$.
	
\begin{figure}[h]
		\centerline{
			\mbox{\includegraphics[width=0.7\textwidth]{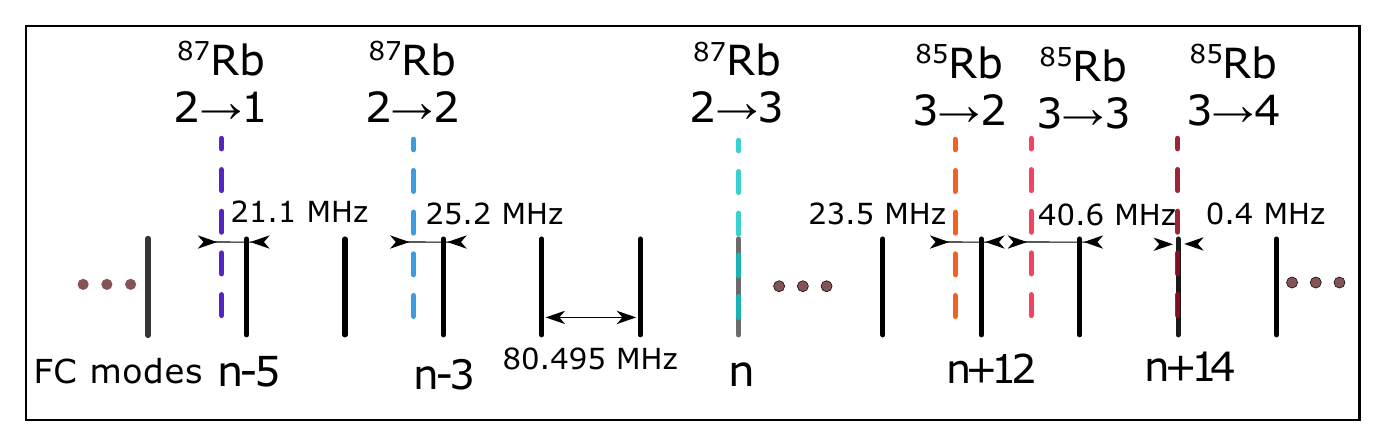}}
		}
		\caption{Relative positions of comb lines within the FC spectrum with respect to the $^{85}$Rb $|5S_{1/2};F=3\rangle \rightarrow |5P_{3/2};F'=2,3,4\rangle$ and $^{87}$Rb $|5S_{1/2};F=2\rangle \rightarrow |5P_{3/2};F'=1,2,3\rangle$ hyperfine transitions for the FC parameters used in the experiment. Note that, when the $n$-th comb mode is resonant with the $^{87}$Rb $|5S_{1/2};F=2\rangle \rightarrow |5P_{3/2};F'=3\rangle$ transition, the $(n+14)$-th comb mode is 0.4 MHz blue detuned from (i.e. almost in resonance with) the $^{85}$Rb $|5S_{1/2};F=3\rangle \rightarrow |5P_{3/2};F'=4\rangle$ transition.   
		}
		\label{Fig2}
\end{figure}
	
The radiation pressure force induced on the $^{85,87}$Rb atoms by the FC, shown in Fig. \ref{Fig1}(b), is the total force obtained by summing the contributions from three hyperfine transitions. 
The force is calculated numerically by solving the optical Bloch equations that describe the excitation of six-level $^{85,87}$Rb atoms by the FC \cite{ban2006} and subsequently using the Ehrenfest theorem. 
More details on the comb force calculation, including the contributions from individual hyperfine transitions, are given in the Supplemental Material. 
The overall agreement of the measured and calculated comb force in Fig. \ref{Fig1} is satisfactory. 
The relative positions of the peaks, as well as the ratios between them are well reproduced for both species. 
The calculated force is however about four times larger than the measured force, and a small broadening of around 1 MHz is observed in the measured peaks. 
This discrepancy can be attributed to several effects that were not taken into account in the theoretical model, such as stray magnetic fields, finite comb mode linewidth, and in particular the spatial beam profile. 
These results are in line with the measured and calculated FC force on $^{87}$Rb atoms in our recent work \cite{santic2019}.
There is also a discrepancy between the calculated and measured force amplitudes around $\delta\approx 0$ which will be clarified in the following paragraphs.
	
In order to understand the relative positions of the comb force peaks in Fig. \ref{Fig1}, it is instructive to study in more detail the hyperfine energy level structure of $^{85,87}$Rb atoms with respect to the FC spectrum. 
Fig. \ref{Fig2} schematically depicts the positions of the comb modes with respect to the three $^{85}$Rb $|5S_{1/2};F=3\rangle \rightarrow |5P_{3/2};F'=2,3,4\rangle$ and the three $^{87}$Rb $|5S_{1/2};F=2\rangle \rightarrow |5P_{3/2};F'=1,2,3\rangle$ hyperfine transitions that are relevant for the atom-comb interaction.
When the $n$-th comb mode is resonant with the $^{87}$Rb $|5S_{1/2};F=2\rangle \rightarrow |5P_{3/2};F'=3\rangle$ transition, the $(n+14)$-th comb mode is 0.4~MHz blue detuned from the $^{85}$Rb $|5S_{1/2};F=3\rangle \rightarrow |5P_{3/2};F'=4\rangle$ transition. 
Taking into account that the natural linewidth of both hyperfine transitions is $\Gamma$=2$\pi \cdot$ 6.07 MHz \cite{steck1, steck2}, the above condition ensures simultaneous excitation of the cooling transitions of both $^{85}$Rb and $^{87}$Rb atoms by the FC.
These two excitations contribute to the FC force peak around $\delta\approx 0$ in Fig. \ref{Fig1}.
The peak at $\delta\approx -25.5$ MHz for $^{87}$Rb is due to the $(n-3)$-rd mode being in resonance with the $|5S_{1/2};F=2\rangle \rightarrow |5P_{3/2};F'=2\rangle$ transition, and the $(n-5)$-th mode in resonance with the $|5S_{1/2};F=2\rangle \rightarrow |5P_{3/2};F'=1\rangle$ transition.
For $^{85}$Rb, the peak at $\delta\approx -23.5$ MHz is due to the $(n+12)$-th comb mode being in resonance with the $|5S_{1/2};F=3\rangle \rightarrow |5P_{3/2};F'=2\rangle$ transition, whereas the peaks at $\delta\approx -40$ MHz and $\delta\approx 40$ MHz are due to the $(n+13)$-th and $(n+12)$-th modes, respectively, being in resonance with the $|5S_{1/2};F=3\rangle \rightarrow |5P_{3/2};F'=3\rangle$ transition.
	
The largest force is obtained for both isotopes for $\delta\approx 0$, where simultaneous excitation of the $^{87}$Rb and $^{85}$Rb cooling transitions is achieved by the $n$-th and $(n+14)$-th comb mode, respectively.
As the $n$-th and $(n+14)$-th comb mode intensities are approximately equal, the larger calculated FC force for $^{85}$Rb compared to $^{87}$Rb reflects the larger transition dipole matrix elements for this transition of $\sqrt{3/2}\cdot3.584\cdot10^{-29}$ C$\cdot$m \cite{steck2}, compared to $\sqrt{7/6}\cdot3.584\cdot 10^{-29}$ C$\cdot$m \cite{steck1} for $^{87}$Rb. 
The actual situation is more complex, as can be seen in the experimental results, where the measured comb force for $\delta\approx 0$ is equal for both species.
Due to the circular polarization of the FC beam and the presence of stray magnetic fields, the $^{85}$Rb $|5S_{1/2};F=3;m_F=+3\rangle \rightarrow |5P_{3/2};F'=4;m_F=+4\rangle$ and  $^{87}$Rb $|5S_{1/2};F=2;m_F=+2\rangle \rightarrow |5P_{3/2};F'=3;m_F=+3\rangle$ stretched transitions dominate all other Zeeman hyperfine transitions.
As these two transitions have equal transition dipole matrix elements of $\sqrt{1/2}\cdot3.584\cdot 10^{-29}$ C$\cdot$m \cite{steck1, steck2}, the resulting comb force is the same for both isotopes. 
This effect could be accounted for in the force calculations by including all Zeeman hyperfine transitions into the optical Bloch equations, but this is rather cumbersome and beyond the scope of this paper.

	
\begin{figure}[h]{\tiny }
		\centerline{
			\mbox{\includegraphics[width=0.7\textwidth]{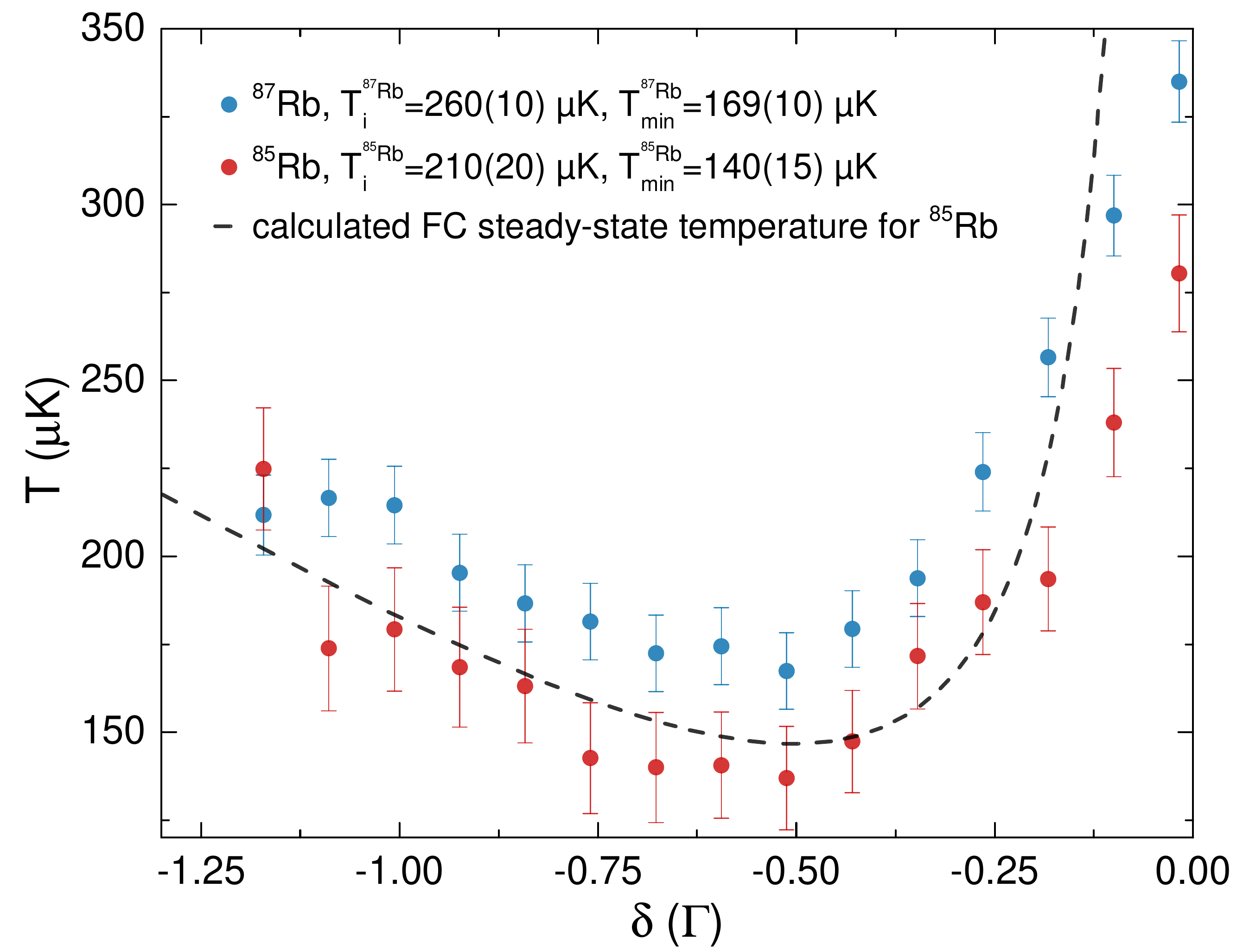}}
		}
		\caption{Temperatures obtained by simultaneous FC cooling of $^{85}$Rb (red circles) and $^{87}$Rb (blue circles) as a function of the comb detuning. $\delta$ denotes the detuning of the $n$-th comb mode from the $^{87}$Rb $|5S_{1/2};F=2\rangle \rightarrow |5P_{3/2};F'=3\rangle$ transition. The initial cloud temperatures are $T_i$=210(20) $\mu$K for $^{85}$Rb and $T_i$=260(10) $\mu$K for $^{87}$Rb, and the FC cooling time is $t_{\text{FC}}=3$~ms. For $^{85}$Rb the steady-state temperature, which approaches the Doppler temperature, is achieved for $\delta\approx-\Gamma/2$. Due to the higher initial cloud temperature for $^{87}$Rb, the steady-state is not achieved during the applied cooling time leading to slightly higher measured temperatures. In general, equal temperature dependence on the comb detuning is obtained for both isotopes. The dashed line indicates the $^{85}$Rb temperature calculated for the relevant experimental parameters using the relation (17) in Ref. \cite{aumiler2012}, i.e. the steady-state temperature obtained for the simple model of 1D FC cooling  of two-level atoms.  		 
		}
		\label{Fig3}
\end{figure}
	
	
FC cooling in one dimension is achieved by using two counter-propagating $\sigma$$^{+}$/$\sigma$$^{-}$-polarized beams. 
We first prepare the pre-cooled clouds of $^{85}$Rb and $^{87}$Rb atoms in a dual-species MOT.
Then we turn off the MOT cooling beams and turn on the FC beams.
The weak CW repumper lasers are left on, as well as the quadrupole magnetic field. 
Since the clouds are in the center of the quadrupole field, the magnetic field is $B\approx0$.
The FC beams are left on for the FC cooling time $t_{\text{FC}}$, after which they are switched off, and the clouds are left to expand freely for several ms.
After a given expansion time, we switch on the MOT cooling beams and image the cloud fluorescence with a camera separately for each species, i.e. the measurement sequence is repeated with the same experimental parameters for each isotope. 
A series of alternating time-of-flight (TOF) images of $^{85}$Rb and $^{87}$Rb clouds are taken in this way at different expansion times.
We then determine the cloud widths by fitting a Gaussian distribution to the spatial distribution of the atoms in the clouds. 
The obtained cloud widths as a function of expansion time $\sigma(t)$ give an accurate measure of the cloud temperature by fitting to the expression $\sigma(t)= \sqrt{\sigma_0^2+\frac{k_{\text{B}}T}{m}t^2}$ \cite{brzozowski2002, yavin2002}, where $\sigma_0$ is the cloud width at $t=0$ (when the FC beams are switched off), $k_{\text{B}}$ is the Boltzmann constant, $m$ is the atomic mass, and $T$ is the cloud temperature.
We repeat the measurement sequence 10 times and average the results to obtain the average temperature and its statistical uncertainty.
Temperatures obtained by FC cooling of $^{85}$Rb and $^{87}$Rb as a function of comb detuning are shown in Fig. \ref{Fig3}.
The total power of the FC beams on the atoms is 20~mW with a beam size (FWHM) of 2.7~mm, resulting in the power and intensity per comb mode of 0.6 $\mu$W and 7.2 $\mu$W/cm$^{2}$, respectively.       
The FC cooling time is $t_{\text{FC}}=3$~ms and TOF images are taken for expansion times between 6-11~ms.       
The initial cloud temperatures of $T_i$ = 210(20) $\mu$K for $^{85}$Rb and $T_i$ = 260(10) $\mu$K for $^{87}$Rb are chosen in order to make the data for the two isotopes clearly distinguishable on the graph. 
For the same cloud initial temperatures $T_i$, the measured FC cooling temperatures for the two isotopes (as a function of comb detuning) would overlap within the measurement error. 
The measured temperatures approach the initial cloud temperatures for large comb detunings.
FC cooling is observed for both isotopes when there is a mode in the comb spectrum that is red detuned from a cooling transition, i.e. when the $n$-th comb mode is red detuned from the $^{87}$Rb $|5S_{1/2};F=2\rangle \rightarrow |5P_{3/2};F'=3\rangle$ transition, and simultaneously the $(n+14)$-th comb mode is red detuned from the $^{85}$Rb $|5S_{1/2};F=3\rangle \rightarrow |5P_{3/2};F'=4\rangle$ transition.
The dashed line in Fig. \ref{Fig3} indicates the steady-state 1D FC cooling temperature for the case of two-level atoms. 
The data is calculated using relation (17) from Ref. \cite{aumiler2012}, and using relevant experimental parameters. 
The following parameters are used: $^{85}$Rb $|5S_{1/2};F=3\rangle \rightarrow |5P_{3/2};F'=4\rangle$ transition dipole matrix element of $\sqrt{3/2}\cdot3.584\cdot 10^{-29}$ C$\cdot$m, $\Gamma$ = 2$\pi \cdot$ 6.07 ~MHz, pulse electric field amplitude $E_0 = 0.9\cdot10^5$~V/m, pulse duration $T_P=300$~fs, pulse area $\theta=\pi/78$, and $f_{\text{rep}}=80.495$~MHz. 
The agreement between the measured $^{85}$Rb temperatures and the 1D FC cooling temperatures calculated for the simple two-level atom model is quite satisfactory. 
In line with the expectations, deviations of the two temperatures are observed due to heating when the comb detuning approaches 0, i.e. when the comb mode approaches resonance.
	
	
\begin{figure}[h]{\tiny }
		\centerline{
			\mbox{\includegraphics[width=0.6\textwidth]{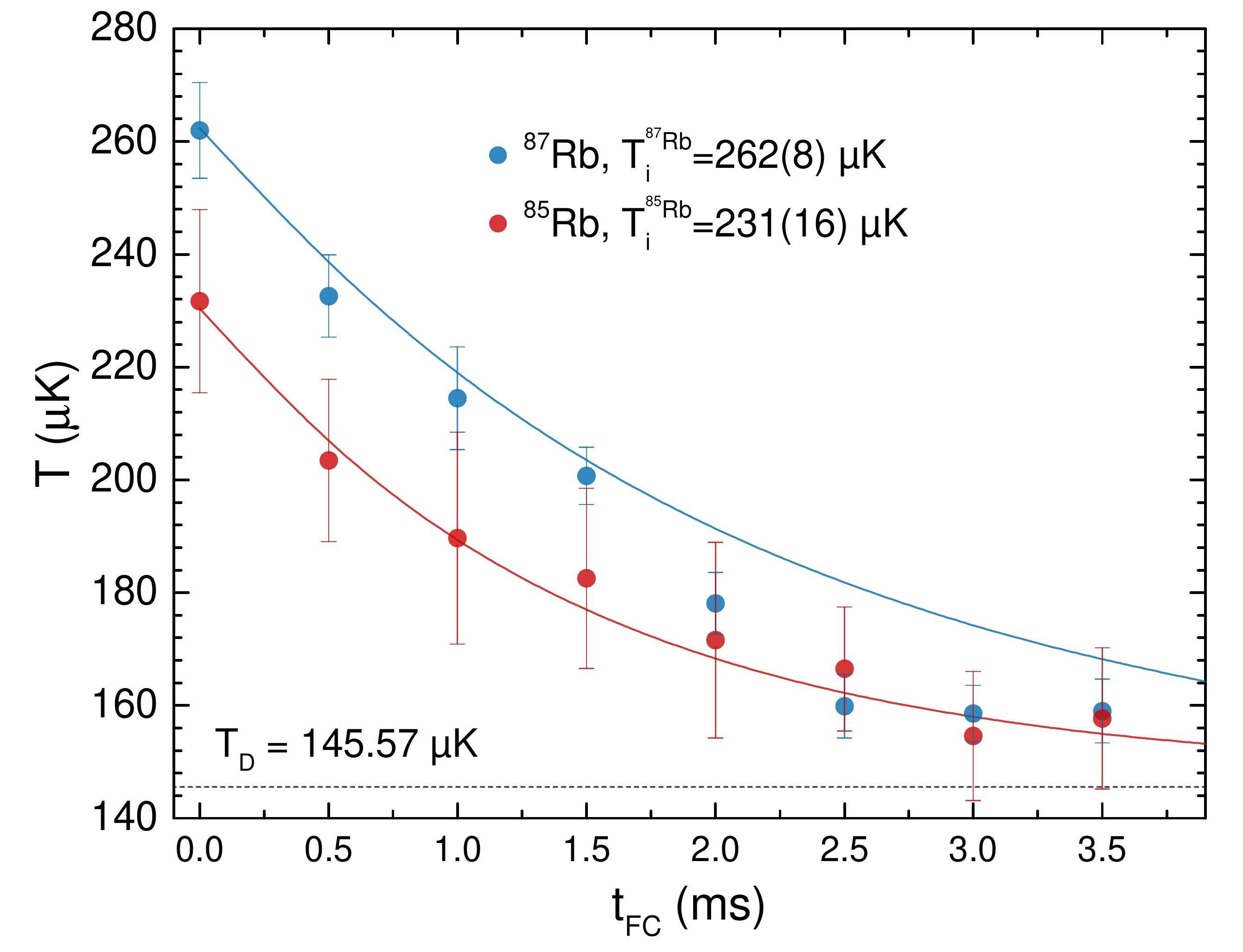}}
		}
		\caption{Temperatures obtained by simultaneous FC cooling of $^{85}$Rb (red circles) and $^{87}$Rb (blue circles) as a function of FC cooling time  $t_{FC}$. The initial cloud temperatures are $T_i$ = 231(16) $\mu$K for $^{85}$Rb and $T_i$ = 262(8) $\mu$K for $^{87}$Rb, with comb detuning $\delta=-\Gamma/2$. As the FC cooling time is increased, the cloud temperature decreases from the initial value $T_i$ prepared in the MOT phase and approaches the Doppler limited steady-state temperature on the time scale of a few ms. Solid lines represent the theoretical estimate for the dependence of the cloud temperature on the FC cooling time (see text for details of the model).		}
		\label{Fig4}
\end{figure}
	
	
The lowest temperature is observed for $\delta\approx-\Gamma/2$, and approaches the Doppler temperature (145.57$~\mu$K \cite{steck2}) for $^{85}$Rb. 
The lowest measured temperature for $^{87}$Rb in Fig. \ref{Fig3} is slightly higher than the Doppler temperature due to the higher initial temperature.
In this case the steady state temperature is not yet reached after the FC cooling time.
This is illustrated in  Fig. \ref{Fig4} where $^{85}$Rb and $^{87}$Rb FC cooling temperatures are shown as a function of the FC cooling time $t_{FC}$.
The initial cloud temperatures are $T_i$ = 231(16) $\mu$K for $^{85}$Rb and $T_i$ = 262(8) $\mu$K for $^{87}$Rb. 
Temperatures are measured for $\delta=-\Gamma/2$, i.e. for the n-th comb mode 3~MHz red detuned from the $^{87}$Rb $|5S_{1/2};F=2\rangle \rightarrow |5P_{3/2};F'=3\rangle$ transition and the $(n+14)$-th comb mode 2.6~MHz red detuned from the $^{85}$Rb $|5S_{1/2};F=3\rangle \rightarrow |5P_{3/2};F'=4\rangle$ transition.
The cloud temperatures decrease from the initial value $T_i$ prepared in the MOT phase, and approach the Doppler limit after a few ms of FC cooling. 
As expected, the steady state temperature is reached sooner when increasing the comb mode intensity and lowering the initial cloud temperature.
	
The solid lines in Fig. \ref{Fig4} represent the theoretical estimates for the dependence of the cloud temperatures on the FC cooling time. 
The estimates are based on a simple model in which the excitation by the comb mode responsible for cooling is considered a CW laser (of the same frequency), and the atomic system is considered a two-level system. 
The model relies on determining the time evolution of the cloud atomic velocity distribution by solving the Fokker-Planck equation for different interaction times, with the radiation pressure force and the diffusion coefficient calculated using standard low-intensity theory for two-level atoms in one dimension (see for example Ref. \cite{metcalf}; more details on the model can be found in Ref. \cite{santic2019}).
For the theoretical results shown in Fig. \ref{Fig4} the following parameters were used: CW laser intensity of 7.2 $\mu$W/cm$^{2}$ (which corresponds to the intensity per comb mode), transition dipole matrix elements $\sqrt{3/2}\cdot3.584\cdot 10^{-29}$ C$\cdot$m and $\sqrt{7/6}\cdot3.584\cdot 10^{-29}$ C$\cdot$m for $^{85}$Rb and $^{87}$Rb, respectively, and $\delta=-\Gamma/2$. 
This simple model well describes the behavior of the measured temperatures, confirming the analogy between FC and CW laser cooling, in line with the results of our previous work \cite{santic2019}. 
The minimum temperature observed in our FC cooling experiment is limited by the low comb mode intensity. 
For higher comb mode intensities, it should be possible to achieve sub-Doppler temperatures as well as FC cooling directly from a room temperature atomic gas \cite{aumiler2012}.
	
In conclusion, we have demonstrated simultaneous cooling of $^{85}$Rb and $^{87}$Rb atoms by using two comb modes from the same FC spectrum. 
Simultaneous cooling of two types of atoms does not affect the cooling of each type, which can be seen from the comparison of FC cooling of $^{87}$Rb in the case of dual-species cooling (this work) and single-species cooling (Ref. \cite{santic2019}).
In addition, we see no evidence that the action of other comb modes within the FC spectrum deteriorates the final FC cooling temperature. 
The minimum observed temperature is limited by the comb mode intensity. 
We confirm the analogy between simultaneous laser cooling of multiple atomic species using a FC and multiple independent CW lasers, in line with Ref. \cite{santic2019}, thus verifying the potential application of the FC for simultaneous multi-channel cooling. 
We believe that the results of this work, together with the advances in novel frequency comb technology that would allow chip-based FCs with significantly increased power per comb line, should contribute to the development of multi-species AIs for space applications. 
In addition, we foresee the application of the results of our work in the experimental investigations of multi-atom interactions and creation of multi-species cold molecules.

\begin{acknowledgments}
The authors acknowledge support from the Croatian Science Foundation (Project Frequency comb cooling of atoms - IP-2018-01-9047).
F. Schmid acknowledges support from the Deutscher Akademischer Austauschdienst (DAAD) and the Croatian Ministry of Science and Education under the German-Croatian bilateral project.
The authors thank T. Udem for reading the manuscript and providing constructive comments.
In addition, the authors acknowledge Neven \v{S}anti\'{c} for early contribution to the development of cold atoms experiment and Ivor Kre\v{s}i\'{c} for the early contribution to the development of theoretical models, as well as Grzegorz Kowzan and Piotr Masłowski for their contribution to the frequency comb stabilization.
\end{acknowledgments}

\section*{SUPPLEMENTAL MATERIAL}

\subsection*{Dual-species magneto-optical trap}

We simultaneously trap and cool $^{85}$Rb and $^{87}$Rb atoms in a dual-species magneto-optical trap (MOT) generated in a stainless steel chamber with a typical vacuum pressure of $\approx$ $10^{-8}$ mbar. 
Standard rubidium dispensers (SAES getters) are used as a source of $^{85}$Rb and $^{87}$Rb atoms in natural abundances. 
A quadrupole magnetic field gradient is generated by a pair of coils in the anti-Helmholtz configuration, with a typical field gradient of $11.3$ G/cm. 
To suppress stray magnetic fields three additional pairs of Helmholtz coils are used.

\begin{figure}[h]
	\centerline{
		\mbox{\includegraphics[width=0.7\textwidth]{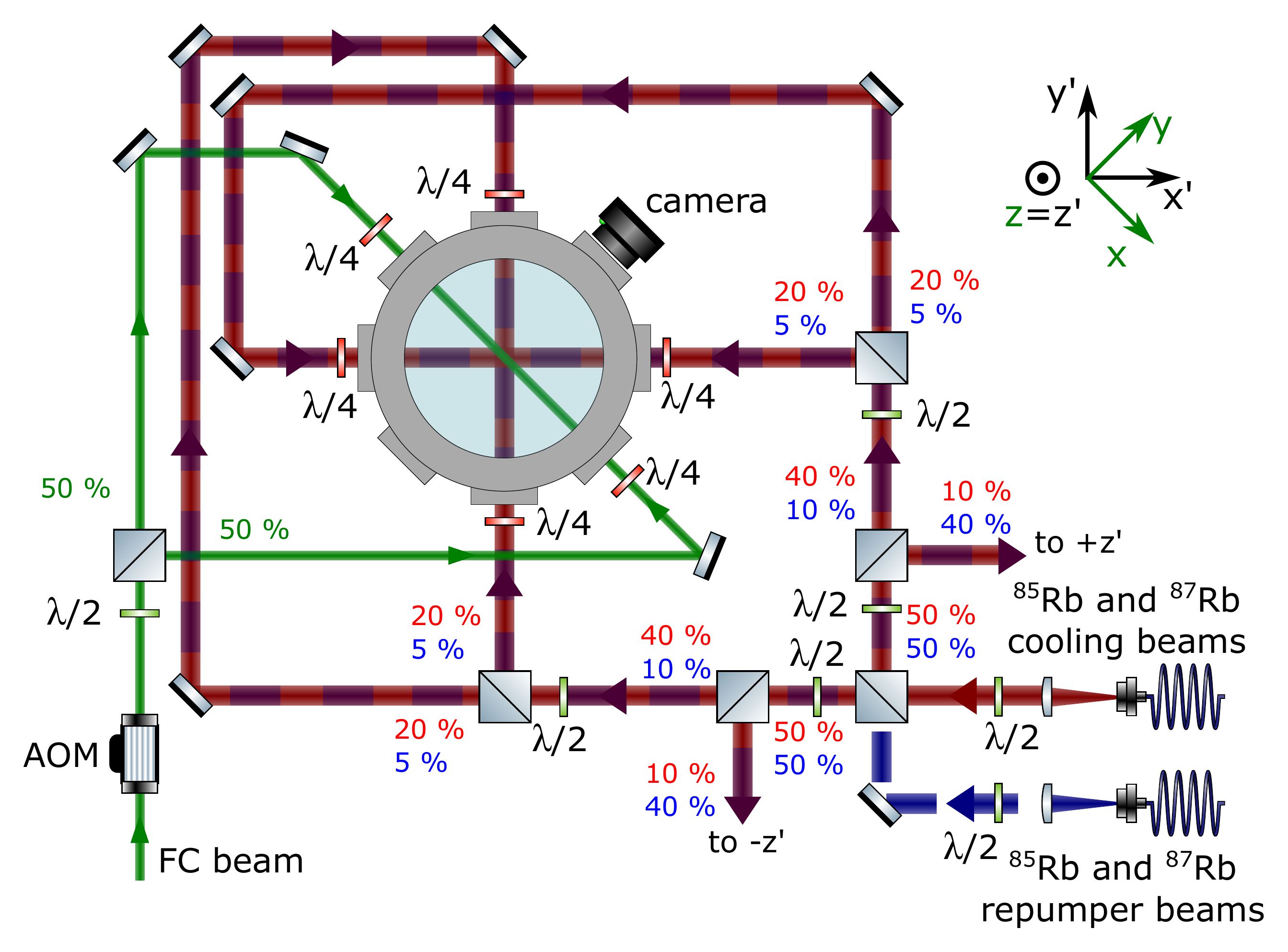}}
	}
	\caption{Schematic diagram of the optical setup used in the experiment. Dual-species MOT relies on four independent frequency-stabilized CW lasers, one cooling and one repumper laser for each species. Cooling and repumper beams are arranged in a standard six-beam configuration, which together with a quadrupole magnetic field (produced by a pair of anti-Helmholtz coils) creates the trapping potential for both species. Anti-Helmholtz coils (not shown in the Figure for clarity) are placed on the opposite sides of the chamber in the xy plane. The cooling laser beams (red) for both species are delivered through a single optical fiber. The repumper laser beams (blue) for both species are delivered with a separate optical fiber. Cooling and repumper laser beams are first combined on a polarizing beam splitter, after which they are divided into six independent beams and directed to the center of the MOT. The power of the beams after each division stage is indicated (in red for cooling lasers and blue for repumper lasers) with respect to the initial power at the fiber outputs. The FC beams, arranged in the counter-propagating configuration, propagate in the horizontal (xy) plane at an angle of 45 degrees to the MOT beams. When the FC radiation pressure force is measured, one FC beam is blocked with a beam stopper. AOM - acousto-optic modulator; $\lambda$/2 - half-wave plate; $\lambda$/4 quarter-wave plate.
	}
	\label{Fig5}
\end{figure}

The frequencies of the two cooling transitions ($|5S_{1/2}; F=3\rangle \rightarrow  |5P_{3/2}; F'=4\rangle$ for $^{85}$Rb and $|5S_{1/2}; F=2\rangle \rightarrow  |5P_{3/2}; F' = 3\rangle$ for $^{87}$Rb) differ by 1126 MHz \cite{steck1, steck2}.
In order to cool and trap both species simultaneously, four independent frequency stabilized continuous-wave (CW) lasers are employed, one cooling and one repumper laser for each species. 
Schematic diagram of the optical setup used in the experiment is shown in Fig. \ref{Fig5}.

The laser system for cooling $^{87}$Rb consists of an external cavity diode laser (ECDL, Moglabs CEL002) which is used as a master laser for injection-locking a solitary laser diode \cite{hadley1986, pagetta2016, hosoya2015}.
This allows accessing the higher output power of a solitary diode compared to an ECDL, while the spectral characteristics are completely determined by the master laser.
The master laser is stabilized to the $^{87}$Rb cooling transition using polarization spectroscopy \cite{harris2006}.
A linewidth (FWHM) of about 100 kHz is measured from a beat note with the frequency comb (FC) \cite{cundiff2005}.

For cooling $^{85}$Rb, an ECDL (Toptica DL100) is used and stabilized to the $^{85}$Rb cooling transition using the Zeeman modulation technique. 
A linewidth (FWHM) of around 600 kHz is obtained from a beat note with the more stable $^{87}$Rb ECDL master laser.
The frequencies of both cooling lasers are controlled using acousto-optic modulators (AOMs).  

Two additional ECDLs (Toptica DL100) are used as repumper lasers.
The laser for repumping $^{87}$Rb atoms from the $|5S_{1/2}; F=1\rangle$ ground hyperfine level is stabilized to the crossover peak arising from the $|5S_{1/2}; F=1\rangle \rightarrow |5P_{3/2}; F'=1\rangle$ and $|5S_{1/2}; F=1\rangle \rightarrow  |5P_{3/2}; F'=2\rangle$ transitions by using direct modulation of the laser current.
The frequency is then shifted to match the $|5S_{1/2}; F=1\rangle \rightarrow |5P_{3/2}; F'= 2\rangle$ transition using an AOM.    
Similarly, the laser for repumping $^{85}$Rb atoms from the $|5S_{1/2}; F=2\rangle$ ground hyperfine level is stabilized to the crossover peak arising from the $|5S_{1/2}; F=2\rangle \rightarrow  |5P_{3/2}; F' = 1\rangle$ and $|5S_{1/2}; F=2\rangle \rightarrow  |5P_{3/2}; F' = 2\rangle$ transitions by using direct modulation of the laser current, and the frequency is then shifted to match the $|5S_{1/2}; F=2\rangle \rightarrow  |5P_{3/2}; F' = 3\rangle$ transition using an AOM.    

To ensure that the $^{85}$Rb and $^{87}$Rb MOTs are well overlapped, the cooling beams for both species are delivered through a single optical fiber. 
The repumping laser beams for both species are delivered with a separate optical fiber. 
The cooling (repumping) laser powers at the fiber outputs are 27(2.8) mW and 40(2.2) mW for $^{85}$Rb and $^{87}$Rb, respectively.
Cooling and repumper laser beams are first combined on a polarizing beam splitter, after which they are divided into six independent beams (three pairs of counter-propagating $\sigma$$^{+}$/$\sigma$$^{-}$ -polarized beams) and directed to the center of the MOT. 
The power of the beams after each division stage is indicated in Fig. \ref{Fig5} (red for cooling lasers and blue for repumper lasers) with respect to the initial power at the fiber outputs.
The beams have a Gaussian spatial profile with a size (FWHM) of 12~mm.  

Loading of the cold atomic clouds and fine-tuning of the cloud temperatures is achieved in two stages. 
In the first stage, atoms are loaded for a duration of 1.3 s. 
The detunings of both cooling lasers are chosen to maximize the number of atoms in the MOT.
In the second stage, with a duration of 0.1 ms, the cooling lasers are tuned closer to the atomic resonance in order to increase the cloud temperatures while keeping the atom numbers constant.
In this way it is possible to simultaneously load $\approx 1\cdot10^6$ $^{85}$Rb atoms and $\approx 3\cdot10^6$ $^{87}$Rb atoms with a temperature that can be fine-tuned in the range of 200-300 $\mu$K.
In typical experimental conditions, we obtain two clouds of cold $^{85}$Rb and $^{87}$Rb atoms with typical 1/e$^{2}$ radii of $\approx $ 0.8 mm, whose centers of mass overlap to within 5 $\%$ of their radii.

\begin{figure}[h]
	\centerline{
		\mbox{\includegraphics[width=0.6\textwidth]{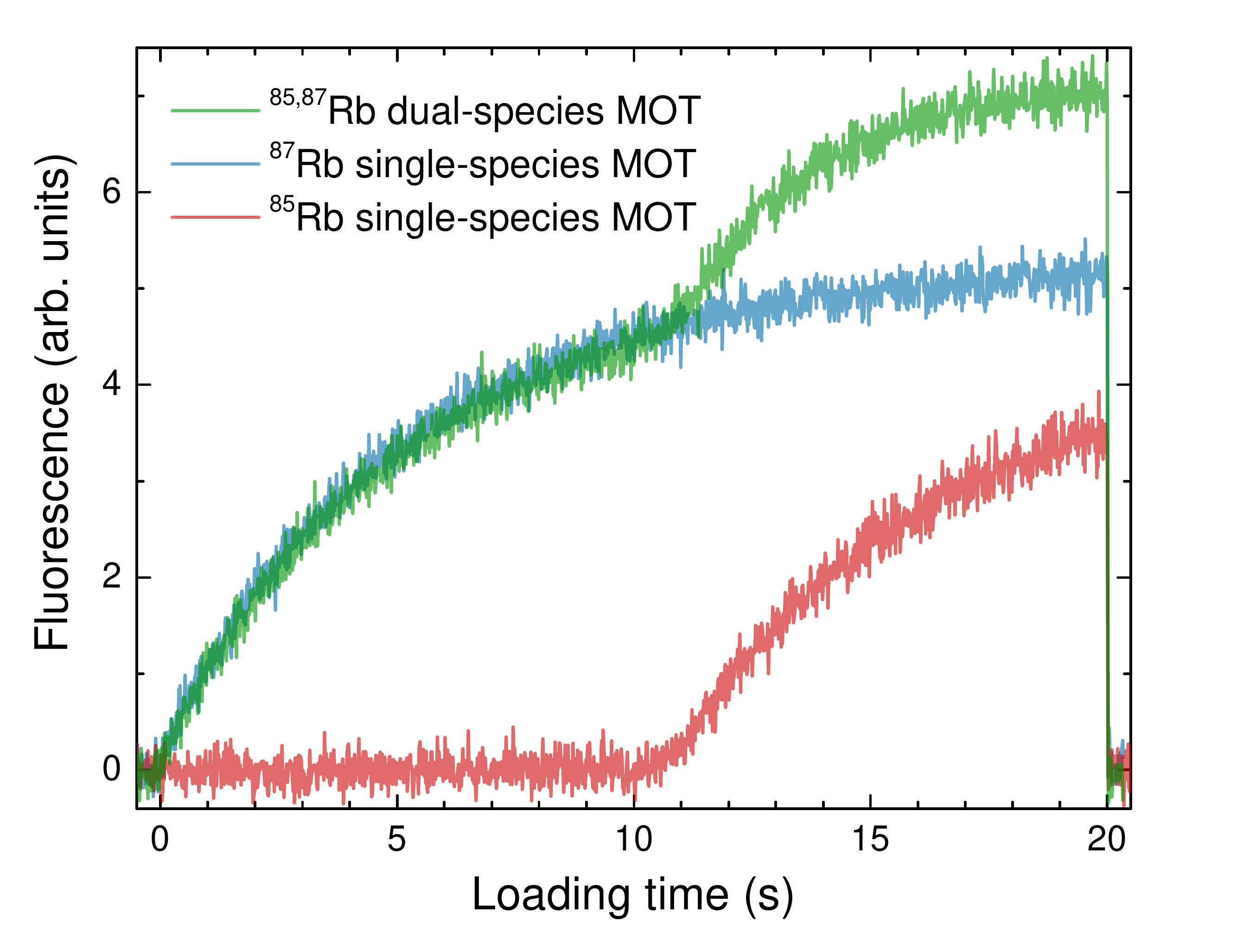}}
	}
	\caption{Loading curves for single-species $^{87}$Rb (blue) and $^{85}$Rb (red) MOTs, and for a dual-species $^{87}$Rb/$^{85}$Rb (green) MOT. For clarity, the $^{85}$Rb cooling and repumping lasers are switched on with a 10 s delay.    
	}
	\label{Fig6}
\end{figure}

In Fig. \ref{Fig6} we show the measured loading curves in the case of single-species $^{87}$Rb (blue) and $^{85}$Rb (red) MOTs, and in the case of a dual-species $^{87}$Rb/$^{85}$Rb (green) MOT.
For clarity, the $^{85}$Rb cooling and repumping lasers are switched on with a 10 s delay.
The loading curves are measured by imaging the fluorescence of the cold clouds with a camera (IDS, UI-3240CP-NIR-GL) and subsequently integrating over the acquired images. 

Note that the total number of trapped atoms in the dual-species MOT (at a loading time of 20 s) is smaller than the sum of the numbers of atoms in the single-species $^{87}$Rb and $^{85}$Rb MOTs.
In the case of the dual-species MOT, trap losses are increased with respect to the single-species MOTs due to atomic collisions between different trapped species.
Similar results were observed in Ref. \cite{suptitz1994}.

\subsection*{Theoretical model for the FC radiation pressure force calculation}

The FC radiation pressure force is calculated numerically by solving the optical Bloch equations that describe the excitation of six-level $^{85,87}$Rb atoms by the FC \cite{ban2006} and subsequently using the Ehrenfest theorem \cite{ilinova2011, aumiler2012}.

The six-level atomic system is comprised of two $5S_{1/2}$ hyperfine ground levels ($F$ = 2, 3 for $^{85}$Rb and $F$ = 1, 2 for $^{87}$Rb) and four $5P_{3/2}$ hyperfine excited levels ($F'$ = 1, 2, 3, 4 for $^{85}$Rb and $F'$ = 0, 1, 2, 3 for $^{87}$Rb).

The pulse train, i.e. the FC, electric field is given by \cite{xu1996}:
\begin{equation}
E_T(t)=\left[\sum_{n=0}^{N} \mathcal{E}(t-nT_R)e^{in\Phi_R} \right] e^{i\omega_L t}, 
\end{equation}
where $N$ is a large integer (order of $10^6$), $\mathcal{E}(t)=E_0\:\mbox{sech}(1.763t/T_p)$ is the single pulse envelope, $T_p$ is the pulse length, $T_R$ is the laser repetition period, $\Phi_R$ is the roundtrip phase acquired by the pulse within the laser cavity, and $\omega_L$ is the central laser frequency. 
The laser spectrum is centered at $\omega_L+\Phi_R/T_R$.
In addition to the pulse train electric field (FC beam), we include a CW laser with amplitude $E_{cw}$ to model the repumper laser which is left on during the interaction of the atoms with the FC.

A system of 21 coupled differential equations for the slowly varying density matrix $\rho_{ij}$ ($i,j$ = 1-6) components is obtained and solved numerically for each species.
The atomic level populations are given by the diagonal density matrix elements, whereas off-diagonal elements represent the coherences.
Due to the presence of CW repumper light, optical pumping to the $F$ = 2 level for $^{85}$Rb and $F$ = 1 level for $^{87}$Rb is prevented and a steady state is achieved after interaction with less than 30 pulses.
To calculate the steady state values of an optical coherence, we average the solution for a pulse train consisting of 150 pulses over the duration of the final 5 pulses.

The force is given by \cite{ilinova2011}:
\begin{equation}
\label{force:th}
F_{FC}=\hbar k \sum_{i,j}\left(\Omega_{i,j}\mbox{Im}(\rho_{ij})\right),
\end{equation}
where $k$ is the wavenumber, $\hbar$ is the reduced Planck constant, $\Omega_{i,j}$ are the effective on-resonance Rabi frequencies, and $\rho_{ij}$ are the optical coherences of the relevant $|5S_{1/2};F=i\rangle \rightarrow |5P_{3/2};F'=j\rangle$ hyperfine transitions.
The effective on-resonance Rabi frequency is given by $\Omega_{ij}=\mu_{ij} E_{\text{comb}}^{\text{eff}} / \hbar$, where $\mu_{ij}$ is the transition dipole matrix element of the $F=i \rightarrow F'=j$ hyperfine transition, and $E_{\text{comb}}^{\text{eff}}$ is the effective electric field amplitude of a single comb mode resonant with the $F=i\rightarrow F'=j $ transition. 
As all modes participating in the comb-atom interactions are close in frequency, the power of the comb modes relevant for the FC force calculation can be taken as equal.
The Rabi frequencies of the relevant transitions differ solely due to different transition dipole matrix elements.

\begin{figure}[h]
	\centerline{
		\mbox{\includegraphics[width=0.6\textwidth]{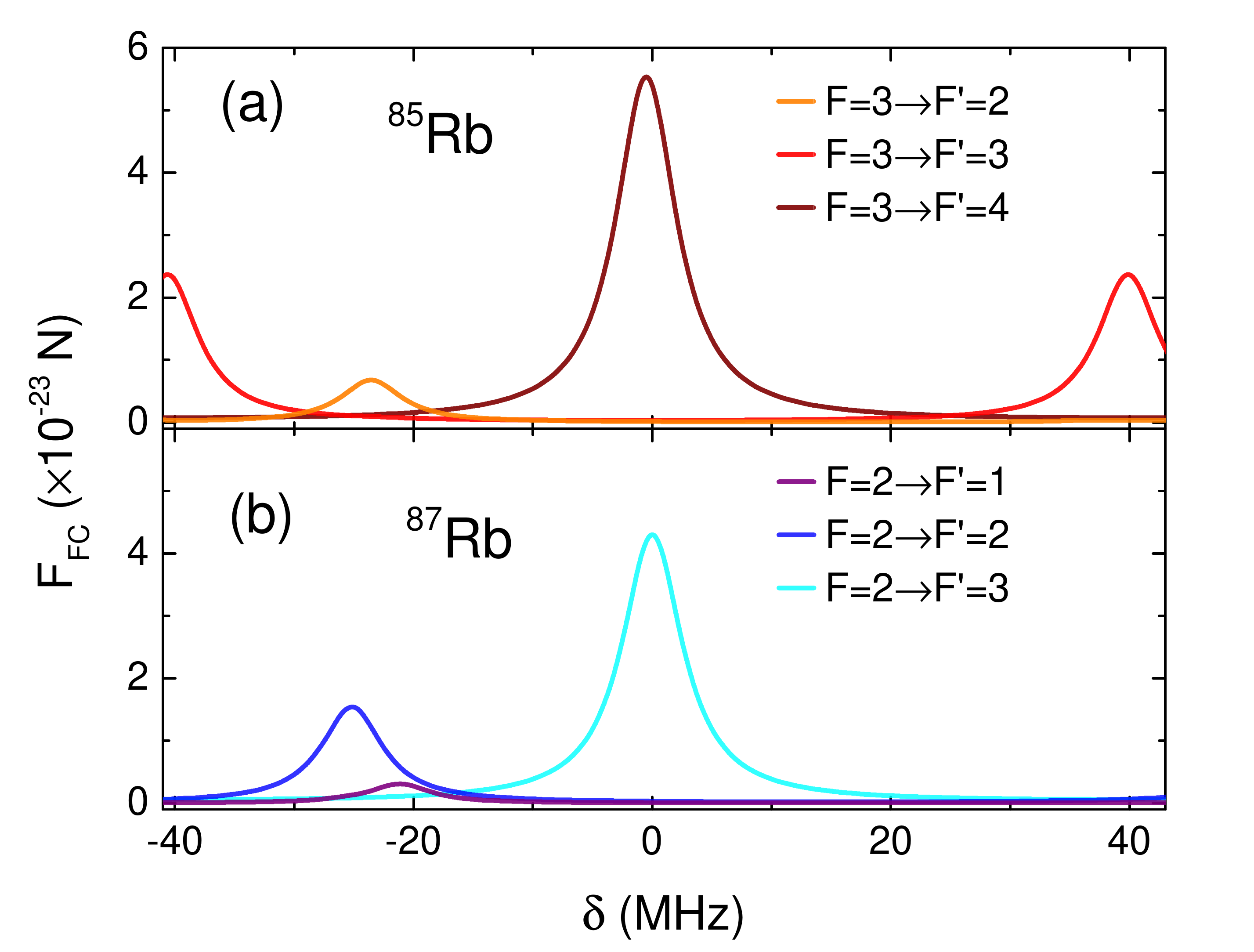}}
	}
	\caption{Calculated FC radiation pressure force as a function of comb detuning $\delta$ for $^{85}$Rb (a) and $^{87}$Rb (b) atoms. $\delta$ denotes detuning of the $n$-th comb mode from the $^{87}$Rb $|5S_{1/2};F=2\rangle \rightarrow |5P_{3/2};F'=3\rangle$ transition. The calculated FC forces include contributions from three $^{85}$Rb $|5S_{1/2};F=3\rangle \rightarrow |5P_{3/2};F'=2,3,4\rangle$ and three $^{87}$Rb $|5S_{1/2};F=2\rangle \rightarrow |5P_{3/2};F'=1,2,3\rangle$ hyperfine transitions. The force contributions are shown separately for each hyperfine transition.    
	}
	\label{Fig7}
\end{figure}

The calculated FC forces presented in Fig. 1(b) in the main text include contributions from three $^{85}$Rb $|5S_{1/2};F=3\rangle \rightarrow |5P_{3/2};F'=2,3,4\rangle$ and three $^{87}$Rb $|5S_{1/2};F=2\rangle \rightarrow |5P_{3/2};F'=1,2,3\rangle$ hyperfine transitions excited by near-resonant comb modes.  
Other non-resonant comb modes do not contribute to the FC force (see e.g. Refs.~\cite{marian2004, cambell2016, santic2019}).

The FC force exerted on $^{87}$Rb atoms arises for the $n$-th comb mode being resonant with the $|5S_{1/2};F=2\rangle \rightarrow |5P_{3/2};F'=3\rangle$ transition, the $(n-3)$-rd comb mode with the $|5S_{1/2};F=2\rangle \rightarrow |5P_{3/2};F'=2\rangle$ transition, and the $(n-5)$-th comb mode with the $|5S_{1/2};F=2\rangle \rightarrow |5P_{3/2};F'=1\rangle$ transition (see Fig. 2 in the main text). 
Following the above, the FC force on a single $^{87}$Rb atom is given by
\begin{equation}
\label{force:th}
F_{FC}= k  E_{\text{comb}}^{\text{eff}} \left(\mu_{23}\mbox{Im}(\rho_{23})+\mu_{22}\mbox{Im}(\rho_{22})+\mu_{21}\mbox{Im}(\rho_{21})\right).
\end{equation}
Here, $\mu_{23}=\sqrt{7/6}\cdot3.584\cdot10^{-29}$~C$\cdot$m,  $\mu_{22}=\sqrt{5/14} \mu_{23}$, and $\mu_{21}=\sqrt{1/14} \mu_{23}$ \cite{steck1}.

The FC force exerted on the $^{85}$Rb atoms arises for the $(n+14)$-th comb mode being resonant with the $|5S_{1/2};F=3\rangle \rightarrow |5P_{3/2};F'=4\rangle$ transition, the $(n+12)$-th comb mode with the $|5S_{1/2};F=3\rangle \rightarrow |5P_{3/2};F'=2\rangle$ transition, and the $(n+13)$-th and $(n+12)$-th comb modes in resonance with the $|5S_{1/2};F=3\rangle \rightarrow |5P_{3/2};F'=3\rangle$ transition (see Fig. 2 in the main text). 
The FC force is given by:
\begin{equation}
\label{force:th}
F_{FC}= k  E_{\text{comb}}^{\text{eff}} \left(\mu_{34}\mbox{Im}(\rho_{34})+\mu_{33}\mbox{Im}(\rho_{33})+\mu_{32}\mbox{Im}(\rho_{32})\right).
\end{equation}
Here, $\mu_{34}=\sqrt{3/2}\cdot3.584\cdot10^{-29}$~C$\cdot$m, $\mu_{33}=\sqrt{35/81} \mu_{34}$, and $\mu_{32}=\sqrt{10/81} \mu_{34}$  \cite{steck2}.

Coupling of the FC to the $F=1$ ($^{87}$Rb) and $F=2$ ($^{85}$Rb) ground levels is neglected because the majority of the population is in the $F=2$ ($^{87}$Rb) and $F=3$ ($^{85}$Rb) levels due to the presence of a repumper laser.
The FC force accelerates the atoms, but their acquired velocity is small, so the Doppler shift and broadening are neglected.

The calculated radiation pressure force induced on the $^{85,87}$Rb atoms due to the FC excitation, shown in Fig. 1(b) in the main text, is the total force obtained by summing the contributions from three hyperfine transitions. 
These contributions are shown in Fig. \ref{Fig7} for $^{85}$Rb (a) and $^{87}$Rb (b) atoms. 
For the simulation the following parameters are used: $E_0=0.9\times 10^5$~V/m, $T_p=300$~fs, $1/T_R=80.495$~MHz and $\Phi_R=0$, $E_{\text{comb}}^{\text{eff}}$ = 5 V/m for the FC beam.
Scanning of the comb spectrum, i.e. changing of the comb detuning $\delta$, is achieved in the simulations by tuning the central laser frequency $\omega_L$ \cite{aumiler2009}. 
For the CW beams a field amplitude of $E_{cw}=40$~V/m is used.

\end{document}